\documentstyle{article}

\begin{document}

\author{{\bf L. Herrera}\thanks{%
On leave of absence from Centro de Astrof\'{\i}sica Te\'{o}rica, Universidad
de Los Andes, M\'{e}rida Venezuela , and Departamento de F\'{\i}sica,
Facultad de Ciencias U.C.V} \ \thanks{%
e-mail: lherrera@gugu.usal.es} \\
Area de F\'{\i}sica Te\'{o}rica{\it , Facultad de Ciencias,}\\
{\it Universidad de Salamanca, }37008, Salamanca, Espa\~{n}a \and {\bf H.
Hern\'{a}ndez\thanks{%
e-mail: hector@ciens.ula.ve}, } \\
Laboratorio de F\'{\i}sica Te\'{o}rica, \\
{\it Departamento de F\'{\i}sica, Facultad de Ciencias, }\\
{\it Universidad de Los Andes, M\'{e}rida 5101, Venezuela} and\\
Centro Nacional de C\'{a}lculo Cient\'{\i}fico \\
Universidad de Los Andes\\
{\it Corporaci\'{o}n Parque Tecnol\'{o}gico de M\'{e}rida, Venezuela} \and 
{\bf L. A. N\'{u}\~{n}ez\thanks{%
e-mail: nunez@ciens.ula.ve}} \\
Centro de Astrof\'{\i}sica Te\'{o}rica{\it , }\\
{\it Departamento de F\'{\i}sica, Facultad de Ciencias,}\\
{\it Universidad de Los Andes, M\'{e}rida 5101, Venezuela} and\\
Centro Nacional de C\'{a}lculo Cient\'{\i}fico\\
Universidad de Los Andes\\
{\it Corporaci\'{o}n Parque Tecnol\'{o}gico de M\'{e}rida, Venezuela} \and %
and {\bf U. Percoco\thanks{%
e-mail: upercoco@ciens.ula.ve}} \\
Centro de Astrof\'{\i}sica Te\'{o}rica, \\
{\it Departamento de F\'{\i}sica, Facultad de Ciencias, }\\
{\it Universidad de Los Andes, M\'{e}rida 5101, Venezuela}}
\date{August 1997}
\title{On the Eccentricity Behaviour \\
of Radiating Slowly Rotating Bodies \\
in General Relativity}
\maketitle

\begin{abstract}
We study different models of radiating slowly rotating bodies up to the
first order in the angular velocity. It is shown that up to this order the
evolution of the eccentricity is highly model-dependent even for very
compact objects.
\end{abstract}

\newpage

\section{Introduction}

In recent papers by Gupta {\it et al }\cite{GuptaEtal96, GuptaEtal97}, it
has been shown that for a specific model of a slowly rotating object (up to
the second order in angular velocity), the ellipticity attains a maximum for
a surface gravitational potential, $\frac MR~\approx ~0.18$ (in geometric
units).

This effect discovered by Chandrasekhar and Miller \cite
{ChandrasekharMiller74}, and initially associated with frame dragging, was
later related to the reversal of the centrifugal force in ultracompact
objects. \cite{AbramowiczPrasanna90, AbramowiczMiller90}.

In this work we study different models of radiating, (up to first order)
slowly rotating objects. The motivation to undertake such a task is
threefold:

\begin{itemize}
\item  firstly, we want to check if the effect reported by Gupta and
collaborators \cite{GuptaEtal96, GuptaEtal97}, appears at first order. In
this case the reversal of the centrifugal force (a second order effect),
although may contribute to, has to be ruled out as the main cause of the
maximum in the eccentricity.

\item  secondly, we would like to clarify if there exists a general tendency
(model independent) in the evolution of eccentricity as the surface
gravitational potential increases (see also \cite{GuptaEtal97} for a
discussion on this point).

\item  finally, it is worth mentioning that the models presented in \cite
{GuptaEtal96, GuptaEtal97} are only valid within the quasi-stationary
approximation. In other words, their results are obtained by identifying the
``history'' of the evolution of the compact object with a sequence of
stationary states. Instead we shall consider here systems which depart from
hydrostatic equilibrium
\end{itemize}

As we shall see below, even for those cases of highly compact matter
configurations, it appears that there is no model independent tendency for
the eccentricity to attain a maximum. In one set of our simulations models
start and end in a stationary state. It is obvious that in these cases the
eccentricity reaches a constant value, but its value depends on the initial
data considered. Several other simulations were preformed using two
different equations of state. The models that emerge never reach stationary
state. In these cases the eccentricity grows up continuously, even for
surface gravitational potential as large as $\frac MR~\approx ~0.31$ .

This paper is organized as follows. In the next section we introduce the
conventions and give a brief description of the method employed to simulate
the evolution of slowly rotating objects. Section 3 presents the three
families of models considered. Finally, in the last section our results are
discussed and contrasted with those obtained by Gupta and collaborators \cite
{GuptaEtal96, GuptaEtal97}.

\section{The Formalism}

Let us consider a nonstatic and axially symmetric distribution of matter
formed by a fluid and radiation. The exterior metric, in radiation
coordinates \cite{Bondi64}, takes the Kerr-Vaidya form \cite{CarmeliCaye77}: 
\begin{eqnarray}
ds^2 &=&\left( 1-\frac{2mr}{r^2+\alpha ^2\cos ^2{\theta }}\right) {\rm d}%
u^2+2{\rm d}u{\rm d}r-2\alpha \sin ^2{\theta }{\rm d}r{\rm d}\phi + 
\nonumber \\
&&+4\alpha \sin \theta ^2\frac{mr}{r^2+\alpha ^2\cos ^2{\theta }}{\rm d}u%
{\rm d}\phi -(r^2+\alpha ^2\cos ^2{\theta }){\rm d}\theta ^2
\label{mexterna} \\
&&-\sin ^2{\theta }\left[ r^2+\alpha ^2+\frac{2mr\alpha ^2\sin ^2{\theta }}{%
r^2+\alpha ^2\cos ^2{\theta }}\right] {\rm d}\phi ^2\;.  \nonumber
\end{eqnarray}
Here, $\alpha $ is the Kerr parameter, representing angular momentum per
unit mass in the weak field limit, and $m$ is the total mass. It is worth
mentioning at this point that the metric above is not a pure radiation
solution and may be interpreted as such only asymptotically \cite
{GonzalezHerreraJimenez79}. A pure rotating radiation solution may be found
in reference \cite{KramerHahner95}. As we shall show below, although the
interpretation of the Carmeli-Kaye metric is not completely clear, the model
dependence of the considered effect is independent of the shape and the
intensity of the emission pulse, and may be put in evidence even for a tiny
radiated energy, $\Delta M_{rad}~=~10^{-12}M(0)$, which for any practical
purpose corresponds to the Kerr metric. The interior metric is written as 
\cite{HerreraJimenez82} 
\begin{eqnarray}
ds^2 &=&e^{2\beta }\left\{ \frac Vr{\rm d}u^2+2{\rm d}u{\rm d}r\right\}
-(r^2+\tilde{\alpha}^2\cos ^2{\theta }){\rm d}\theta ^2  \nonumber \\
&&+2\tilde{\alpha}e^{2\beta }\sin ^2{\theta }\left\{ 1-\frac Vr\right\} {\rm %
d}u{\rm d}\phi -2e^{2\beta }\tilde{\alpha}\sin ^2{\theta }{\rm d}r{\rm d}\phi
\label{minterna} \\
&&-\sin ^2{\theta }\left\{ r^2+\tilde{\alpha}^2+2\tilde{\alpha}^2\sin ^2{%
\theta }\frac Vr\right\} {\rm d}\phi ^2\;.  \nonumber
\end{eqnarray}

In equations (\ref{mexterna}) and (\ref{minterna}), $u~=~x^0$ is a time like
coordinate, $r~=~x^1$ is the null coordinate and $\theta ~=~x^2$ and $\phi
~=~x^3$ are the usual angle coordinates. The $u$-coordinate is the retarded
time in flat space-time and, therefore, $u$-constant surfaces are null cones
open to the future.

The Kerr parameter for the interior space-time (\ref{minterna}) is denoted $%
\tilde{\alpha}$ and, for the present work and for sake of simplicity, will
be considered constant and relevant only (as well as $\alpha $ in eq. (\ref
{mexterna})) up to the {\it first order}. Notice that in these coordinates
the $r~=~constant~=~r_s$ surfaces are not spheres but{\it \ oblate spheroids}%
, whose eccentricity depends upon the Kerr parameter $\tilde{\alpha}$ and is
given by 
\begin{equation}
e^2=1-\frac{r_s^2}{r_s^2+\tilde{\alpha}^2}\;,  \label{eccentricity}
\end{equation}
with $r_s$ representing the shell where the eccentricity is evaluated.
Observe that eccentricity as defined by (\ref{eccentricity}), although it
yields the correct Newtonian limit and is the natural definition in the
context of metrics (\ref{mexterna}) and (\ref{minterna}), is not invariantly
defined . However the main argument to use it here, stems from the fact that
it is that parameter the one whose evolution is studied in references \cite
{GuptaEtal96, GuptaEtal97}.

The metric elements $\beta $ and $V$ in eq. (\ref{minterna}), are functions
of $u$, $r$ and $\theta $. A function $\tilde{m}(u,r,\theta )$ can be
defined by 
\begin{equation}
V=e^{2\beta }\left( r-\frac{2\tilde{m}(u,r,\theta )r^2}{r^2+\tilde{\alpha}%
^2\cos ^2{\theta }}\right) \;,  \label{eq_massin}
\end{equation}
which is the generalization, inside the distribution, of the ``mass aspect''
defined by Bondi and collaborators \cite{BondiVandenburgMetzner62} and in
the static limit coincides with the Schwarzschild mass.

In this work the modeling is performed by means of a general method
presented a few years ago \cite{HerreraEtal94}, which allows the generation
of axially symmetric slowly rotating (up to the first order), radiating
solutions from known static ``seed'' solutions.

Only a very brief description of the method is given here, we refer the
reader to \cite{HerreraEtal94} for details.

The method starts by defining two auxiliary functions 
\begin{equation}
\tilde{\rho}=\frac{\rho -P\omega _x}{1+\omega _x}  \label{ro_efec}
\end{equation}
and 
\begin{equation}
\tilde{P}=\frac{P-\rho \omega _x}{1+\omega _x}\;.  \label{pres_efec}
\end{equation}
which are called the {\it effective variables}. In equations (\ref{ro_efec})
and (\ref{pres_efec}) $\omega _x$ represents the radial component of the
velocity of a fluid element as measured by a local minkowskian observer.
Also in the above equations the physical (local) pressure and density are
denoted by $\rho $ and $P$.

Next, it can be easily shown using the field equations that the metric
elements $\tilde{m}$ and $\beta $ can be expressed as 
\begin{equation}
\tilde{m}=\int_0^r{\rm d}\bar{r}4\pi \bar{r}^2\tilde{\rho}  \label{masa}
\end{equation}
and 
\begin{equation}
\beta =\int_r^{a(u)}\frac{2\pi \bar{r}^2{\rm d}\bar{r}}{\bar{r}-2\tilde{m}}%
\left( \tilde{\rho}+\tilde{P}\right) \;.  \label{beta}
\end{equation}
where $r=$ $a(u)$ is the equation of the boundary surface.

It is clear that, if the radial dependence of $\tilde{\rho}$ and $\tilde{P}$
are borrowed from the ``seed'' static solution, the metric elements $\tilde{m%
}$ and $\beta $ , can be determined from eq. (\ref{masa}) and (\ref{beta}),
up to some functions of the time-like coordinate $u.$ In the context of this
approach the radial dependence of $\tilde{\rho}$ and $\tilde{P}$ is assumed
to be the same as in the static ``seed'' solution.

The rationale behind the assumption on the $r$ dependence of the{\it \
effective variables }$\tilde{P}$ and $\tilde{\rho}$, can be grasped in terms
of the characteristic times for different processes involved in a collapse
scenario. If the hydrostatic time scale ${\cal T}_{HYDR}$, which is of the
order $\sim 1/\sqrt{G\rho }$ (where $G$ is the gravitational constant and $%
\rho $ denotes the mean density) is much smaller than the {\it %
Kelvin-Helmholtz} time scale ( ${\cal T}_{KH}$ ), then in a first
approximation the inertial terms in the equation of motion can be ignored 
\cite{KippenhahnWeigert90} . Therefore in this first approximation
(quasi-stationary approximation) the $r$ dependence of $P$ and $\rho $ are
the same as in the static solution. Then the assumption that the{\it \
effective variables} (\ref{ro_efec}) and (\ref{pres_efec}) have the same $r$
dependence as the physical variables of the static situation, represents a
correction to that approximation, and is expected to yield good results
whenever ${\cal T}_{KH}\gg {\cal T}_{HYDR}$. Fortunately enough, ${\cal T}%
_{KH}\gg {\cal T}_{HYDR}$, for almost all kind of stellar objects.

Those functions of the time-like coordinate $u$ that remain arbitrary can be
obtained from a system of ordinary differential equations ({\it The System
of Surface Equations}) emerging from the field equations evaluated at the
boundary surface and the junction conditions.

The equations corresponding to the junction conditions (eq. (49), and (50)
in \cite{HerreraEtal94}) lead to 
\begin{equation}
\beta _{1a}\left( 1+\dot{a}\right) =f(u)\alpha  \label{definiefe}
\end{equation}
and 
\begin{equation}
\tilde{m}_{1a}\left( 1+\dot{a}\right) =g(u)\alpha \;.  \label{definige}
\end{equation}
Where $f(u)$ and $g(u)$ are arbitrary functions of their arguments,
differentiation with respect to $u$ and $r$ are denoted by subscripts $0$
and $1$, respectively and the subscript $a$ indicates that the quantity is
evaluated at the boundary surface. Finally, in radiation coordinates, $\dot{a%
}$ takes the form 
\begin{equation}
\dot{a}=\frac{{\rm d}a}{{\rm d}u}=\frac{\omega _{xa}}{1-\omega _{xa}}\left(
1-\frac{2\tilde{m}_a}a\right) \;.  \label{a_punto2}
\end{equation}
Also, from the junctions conditions (eq. (40) in \cite{HerreraEtal94}) we
get, 
\begin{equation}
\beta _{1a}\left( 1-\frac{2\tilde{m}_a}a\right) -\beta _{0a}=\frac{\tilde{m}%
_{1a}}{2a}\quad \ ,  \label{segforma}
\end{equation}
or equivalently 
\begin{equation}
2a\beta _{1a}\left( 1+\dot{a}-\frac{2\tilde{m}_a}a\right) =\tilde{m}_{1a}
\label{segforma2}
\end{equation}

From equations (\ref{segforma2}), (\ref{definiefe}) and (\ref{definige}) we
obtain an expression relating $f(u)$ and $g(u),$ namely 
\begin{equation}
2a\left( 1+\dot{a}-\frac{2\tilde{m}_{a}}{a}\right) f(u)=g(u)\ .
\label{relaefege}
\end{equation}
However, it can be checked by simple inspection that neither the field
equations nor the junction conditions impose further restrictions on these
functions of $u$, therefore one of them remains completely arbitrary for
each model.

Expanding (\ref{eccentricity}) for $\tilde{\alpha}\ll 1$ , we get 
\begin{equation}
e=\frac 1{r_s}\tilde{\alpha}-\frac 12\frac 1{r_s^3}\tilde{\alpha}^3+\cdots \
,  \label{expeccen}
\end{equation}
thus, as expected, up to first order the eccentricity is proportional to $%
\tilde{\alpha}.$ Now equations (\ref{definige}) and (\ref{relaefege}) lead
to 
\begin{equation}
\tilde{\alpha}=\frac{\tilde{m}_{1a}\left( 1+\dot{a}\right) }{2a\left( 1+\dot{%
a}-\frac{2\tilde{m}_a}a\right) f(u)}  \label{alpha}
\end{equation}
or, using eq. (\ref{masa}), 
\begin{equation}
\tilde{\alpha}=2\frac{\pi a\tilde{\rho}_a(u)\left( 1+\dot{a}\right) }{\left(
1+\dot{a}-\frac{2\tilde{m}_a}a\right) f(u)}\ .  \label{alpharho}
\end{equation}
It is easy to see that two different types of variables are needed to
determine $\tilde{\alpha}$ from (\ref{alpharho}). On one hand, we notice $a,%
\dot{a},\tilde{\rho}_a,$ and $\tilde{m}_a$ which can be obtained from the
integration of the {\it Surface Equations} and the particular ``seed''
solution selected to be modeled. On the other hand, as we have stated
before, $f(u)$ remains completely arbitrary and its choice completes the
characterization of the model.

In the next section we shall present three different families of models
obtained from different static equations of state.

\section{The Modeling}

In order to show that there is no model independent tendency for the
eccentricity to attain a maximum, we shall consider three different families
of equations of state. Each one will be modeled with three different choices
of the arbitrary function $f(u)$. We shall work out three models, previously
studied for the spherical (nonrotating) case: {\it Schwarzschild-like}\cite
{Tolman39, HerreraJimenezRuggeri80},{\it Tolman-VI-like }\cite{Tolman39,
HerreraJimenezRuggeri80} and {\it Tolman-V-like }\cite{Tolman39,
PatinoRago83, AguirreHernandezNunez94} solutions. In the static limit the 
{\it Schwarzschild-like} homogeneous solution represents an incompressible
fluid of constant density. Static {\it Tolman VI }solution approaches that
of a highly relativistic Fermi Gas and, therefore, with the corresponding
adiabatic exponent of 4/3. And, for the static{\it \ Tolman V }solution, the
radiation relation $P/{\rho }\sim 1/3$, is maintained at the center of the
distribution during the contraction.

All the above equations of state will be used as a ``seed'' solution to
study the evolution of the eccentricity. The first family to be considered
is the Schwarzschild solution. It is the same example presented in ref. \cite
{HerreraEtal94}. 
\begin{equation}
\tilde{\rho}=k(u)=\frac 3{8\pi }\frac{\left( 1-F(u)\right) \ }{A(u)\ ^2}\;,
\label{shwroef}
\end{equation}
and 
\begin{equation}
\tilde{\alpha}=\frac{3\left( 1-\,F(u)\right) \left( 1+F(u)\left( \Omega
(u)-1\right) \right) }{4\,A(u)\ F(u)\ \Omega (u)\ f(u)}\ ;  \label{shwalpha}
\end{equation}
where the exterior radius $a$, the total mass $m$, and the timelike
coordinate $u$ are scaled by the initial total mass $m(u=0)=m(0)$, i.e. $%
A=a/m(0)$; $\;M=m/m(0)$ and $u=u/m(0)$. We have also defined 
\begin{equation}
F(u)=1-\frac{2M(u)}{A(u)}  \label{eFe}
\end{equation}
and 
\begin{equation}
\Omega (u)=\frac 1{1-\omega _{xa}}\ .  \label{Omega}
\end{equation}
The second case is the slowly rotating {\it Tolman-V-like} models. In this
case we have 
\begin{equation}
\tilde{\rho}=\frac 1{8\pi }\left( \frac{\delta (u)}{r^2}+z(u)r^{\frac
13}\right)  \label{to5roef}
\end{equation}
where 
\begin{equation}
\tilde{\alpha}=\frac{\left( 1-F(u)\right) \left( 1+F(u)\left( \Omega
(u)-1\right) \right) }{3\ \,A(u)\ F(u)\ f(u)}\ .  \label{to5alpha}
\end{equation}
with 
\begin{equation}
\delta (u)=\frac 27\left( F(u)-1\right) \left( 2\,\Omega (u)-5\right)
\label{aux1to5}
\end{equation}
and 
\begin{equation}
z(u)=\frac{10}{21\,A(u)^{7/3}}\left( 1-F(u)\,\right) \left( 4\,\Omega
(u)-3\right) .  \label{aux2to5}
\end{equation}

Finally we take the {\it Tolman VI-like} family of models, with the
effective density represented by

\begin{equation}
\tilde{\rho}=\frac{3h(u)}{r^2}=\frac 1{8\pi }\frac{\left( 1-F(u)\right) }{r^2%
}  \label{to6roef}
\end{equation}
and for the eccentricity we have found 
\begin{equation}
\tilde{\alpha}=\frac{\left( 1-F(u)\right) \left( 1+F(u)\left( \Omega
(u)-1\right) \right) }{4\,A(u)\ F(u)\ \Omega (u)\ f(u)}.  \label{to6alpha}
\end{equation}

As we have mentioned above, the evolution of these three families of
eccentricities (\ref{shwalpha}), (\ref{to6alpha}), (\ref{to5alpha}) have
been calculated for three different expressions of $f(u)$, namely: 
\begin{equation}
f(u)=\left\{ 
\begin{tabular}{l}
$cte=1$ \\ 
$1+u$ \\ 
$1+C\exp -\left( \frac{u-u_{trig}}\sigma \right) ^2$%
\end{tabular}
\right.  \label{formefe}
\end{equation}
with $u_{trig}$ a triggering parameter, $C$ a modulation constant and $%
\sigma $ the width of the pulse.

Simulations have been carried out for a large number of sets of initial data
and emission pulses. We shall present in the next section only the most
relevant cases.

The radiation flux has been assumed to be in the {\it free streaming out
limit }approximation, i.e. neutrinos and/or photons mean free path is of the
order of the dimensions of the spheroid. It is worth mentioning that for the
modeling performed, the shape and the intensity of the emission pulse does
not seem to play an important role, we shall only exhibit the results
obtained with a Gaussian pulse centered at $u=u_p$ 
\begin{equation}
-\dot{M}=L=\frac{\Delta M_{rad}}{\lambda \sqrt{2\pi }}\exp \frac 12\left( 
\frac{u-u_p}\lambda \right) ^2,  \label{eq:mpunto}
\end{equation}
with $\lambda $ the width of the pulse and $\Delta M_{rad}$ the total mass
lost in the process. All models presented in the next section have been
simulated using 
\begin{equation}
\Delta M_{rad}=0.001\qquad \lambda =0.1\qquad u_p=15  \label{pulset}
\end{equation}

The sets of initial data relevant to be discussed are:

\begin{itemize}
\item  {\it Schwarzschild-like} (displayed in Figures 1 and 2) 
\begin{equation}
\frac{M(0)}{A(0)}=0.185;\quad \Omega (0)=0.933;\quad A(0)=5.404;\quad
F(0)=0.630  \label{shwset1}
\end{equation}
and 
\begin{equation}
\frac{M(0)}{A(0)}=0.303;\quad \Omega (0)=0.933;\quad A(0)=3.300;\quad
F(0)=0.394  \label{shwset2}
\end{equation}

\item  {\it Tolman V-like} (displayed in Figure 3 ) 
\begin{equation}
\frac{M(0)}{A(0)}=0.217;\quad \Omega (0)=0.809;\quad A(0)=4.600;\quad
F(0)=0.565  \label{to5set}
\end{equation}

\item  {\it Tolman VI-like} (displayed in Figure 4 ) 
\begin{equation}
\frac{M(0)}{A(0)}=0.270;\quad \Omega (0)=1.000;\quad A(0)=3.700;\quad
F(0)=0.459  \label{to6set}
\end{equation}
\end{itemize}

The running time for the models (values for $u$) are controlled by the
physical relevance of the variables involved. As we shall see in the next
section the behavior of $\tilde{\alpha}$ is highly model dependent, no
matter how compact the object becomes.

\section{Discussion of Results}

Let us start by analyzing the {\it Schwarzschild-like} model. Figures 1a
through 1c display the evolution of the eccentricity for the above mentioned
choices of $f(u)$ in equation (\ref{formefe}) with the set of initial data (%
\ref{shwset1}). Figure 1d indicates the evolution of $F(u)$. Observe that
cases 
\begin{equation}
f(u)=\left\{ 
\begin{tabular}{l}
$cte=1$ \\ 
$1+C\exp -\left( \frac{u-u_{trig}}\sigma \right) ^2$%
\end{tabular}
\right.  \label{shwcase}
\end{equation}
yield results very similar to those reported in \cite{GuptaEtal96} although
we are considering only first order terms ! These result are presented in
Figures 1a and 1c, respectively. From Figures 1a, 1c and 1d we can see that
the eccentricity for the above two cases present a maximum when the surface
gravitational potential reaches 
\begin{equation}
\frac{M(u)}{A(u)}=\frac 12\left( 1-F(u)\right) \approx 0.18\ .
\label{shwma1}
\end{equation}
The other model, corresponding to the case $f(u)=1+u$ also gives a
stationary value for the eccentricity for the above value for the
gravitational potential at the surface of the configuration.

Thus, up to the first order this model would seem to confirm the result
found in \cite{GuptaEtal96, GuptaEtal97}. However, this is not so. Indeed,
Figures 2a to 2d represent the evolution of the eccentricity, $e(u)$ , and $%
F(u)$ , for another set of initial data (see equations (\ref{shwset2})). Now
the stationary value for $e(u)$ is obtained for $\frac{M(u)}{A(u)}~\approx
~0.33$ . It is worth noticing that for a given set of initial data the final
value of $e(u)$ is independent of the $f(u)$ .

Let us now turn to consider the {\it Tolman V-like} family of models with
the set of initial data (\ref{to5set}). Figures 3a through 3c exhibit the
evolution of the eccentricity concerning the functions $f(u)$ in equation (%
\ref{formefe}). In Figure 3d the evolution of $F(u)$ is displayed. As it is
apparent from these figures, there is no maximum in the value of the
eccentricity, even though the surface gravitational potential goes well
beyond the supposed critical value of $\frac{M(u)}{A(u)}\approx 0.18$ (see
Figure 3d).

Finally, Figures 4a to 4d refer to {\it Tolman VI-like} family of models,
and point in the same direction as the previous case. In fact, Figures 4a to
4c show that either the eccentricity grows up continuously ($f(u)$ is one of
the above eq. (\ref{shwcase})) or attains a minimum and grows up again ($%
f(u)~=~1+u$), even though the surface gravitational potential may reach
values as high as $\frac{M(u)}{A(u)}~\approx ~0.31$ (Figure 4d).

Summarizing our results. It is clear that, up to the first order in angular
parameters, the evolution of the eccentricity is highly model dependent. If
second order contributions are going to be considered, they should be
assumed much smaller than the first order ones. Therefore, it is not
reasonable to expect that second order contributions could bend the curves
displayed in Figures 3a, 3c, 4a and 4c as to create a maximum for $e(u)$ or
that they can subdue the dependence of the final value of $e(u)$ on the
initial data for the {\it Schwarzschild-like} models. In other words, we can
not see how the second order effects might affect qualitatively the model
dependence exhibited here.

Of course, it remains to be seen whether the effects described in \cite
{GuptaEtal96, GuptaEtal97} appear in a model independent way in the exact
theory of relativistic rotating bodies. For that purpose, however, we would
need an exact interior solutions for the axially symmetric source, which, as
it is well known, is not available yet.

We would like to conclude with the following comment. Although the main
purpose here has been to bring out the model dependence of the evolution of
the eccentricity, and not specific modeling of rotating sources, it should
be noticed that in the examples above the energy density is always positive
and larger than pressure everywhere within the fluid distribution and the
fluid velocity as measured by the locally Minkowskian observer is always
less than one. Although we do not include figures of physical variables
(some of them for the case $f=1$ may be found in reference \cite
{HerreraEtal94}), the reader may convince himself of that by noticing that
the three models considered here have been studied in detail in the
spherically symmetric case \cite{HerreraJimenezRuggeri80, PatinoRago83,
AguirreHernandezNunez94}, in which all models have a correct physical
behaviour. Of course we do not expect that first order perturbation may
change that situation.

Two of us (H.H. and L.A.N) gratefully acknowledge the financial support by
the Consejo de Desarrollo Cient\'{\i}fico Human\'{\i}stico y Tecnol\'{o}gico
de la Universidad de Los Andes, under project C-720-95-B and by the Programa
de Formaci\'{o}n e Intercambio Cient\'{\i}fico (Plan II).

\newpage\ 

\begin{center}
{\bf Figure Captions}
\end{center}

\begin{description}
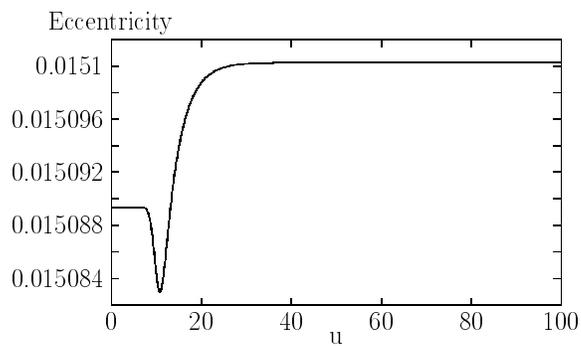
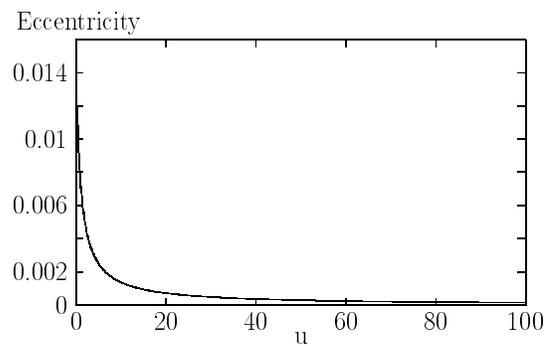
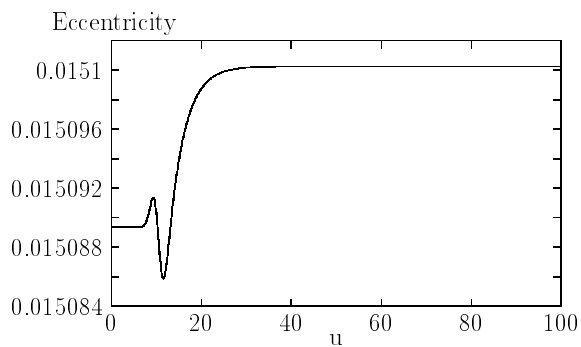
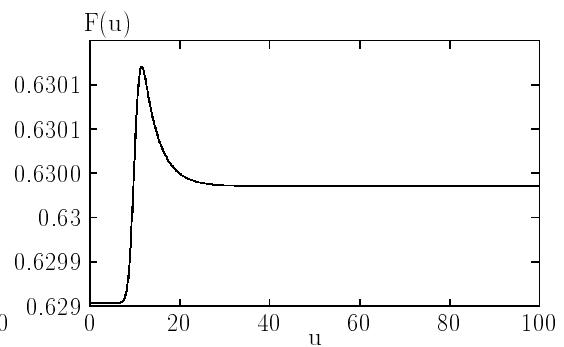
\item[Figure 1]  The evolution of the eccentricity for the {\it %
Schwarzschild-like} model with $\frac{M(0)}{A(0)}=0.185$. Figures 1a through
1c correspond to functions: $f(u)~=~cte=1;$ $f(u)~=~1+u;$ and $%
f(u)~=~1+C\exp -\left( \frac{u-u_{trig}}\sigma \right) ^2$ , respectively.
Figure 1d indicates the evolution of the gravitational potential at the
surface: $F(u)=1-\frac{2M(u)}{A(u)}$ .

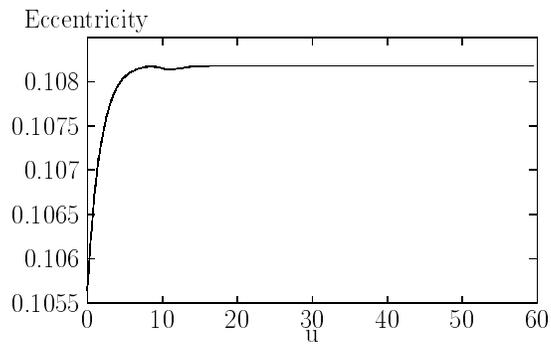
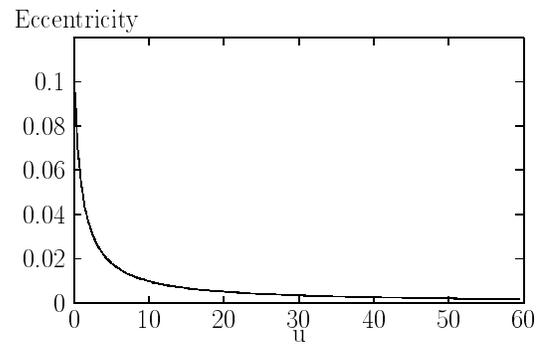
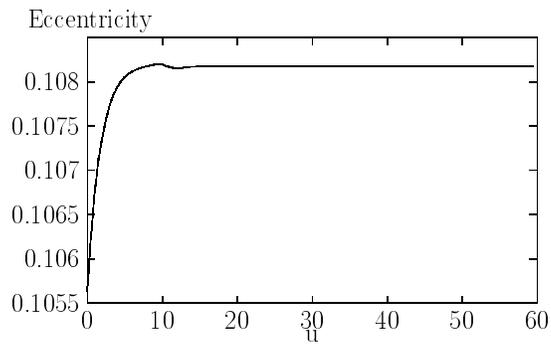
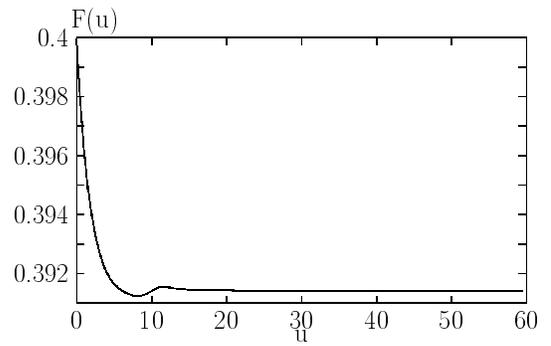
\item[Figure 2]  The evolution of the eccentricity for the {\it %
Schwarzschild-like} model with $\frac{M(0)}{A(0)}=0.303.$ Figures 2a through
2c correspond to functions: $f(u)~=~cte=~1;$ $f(u)~=~1+u;$ and $%
f(u)~=~1+C\exp -\left( \frac{u-u_{trig}}\sigma \right) ^2$ , respectively.
Figure 2d indicates the evolution of the gravitational potential at the
surface: $F(u)=1-\frac{2M(u)}{A(u)}$ .

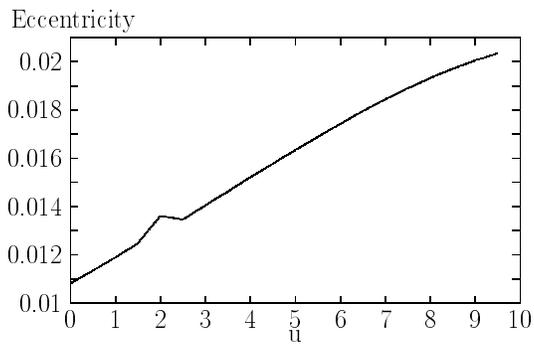
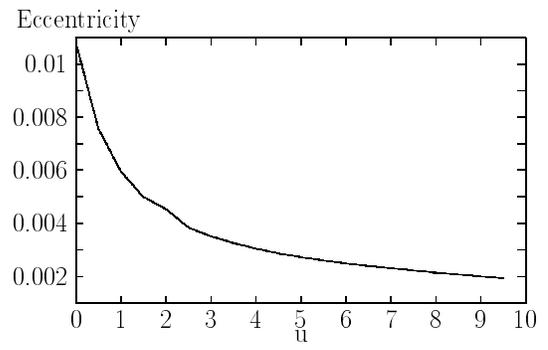
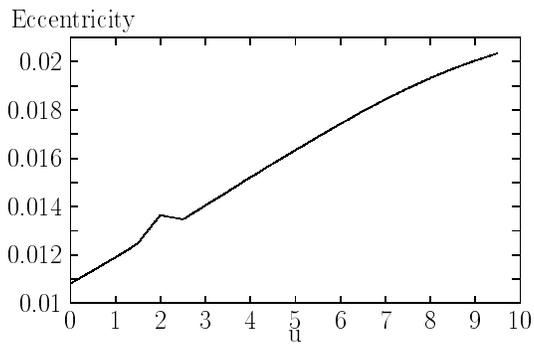
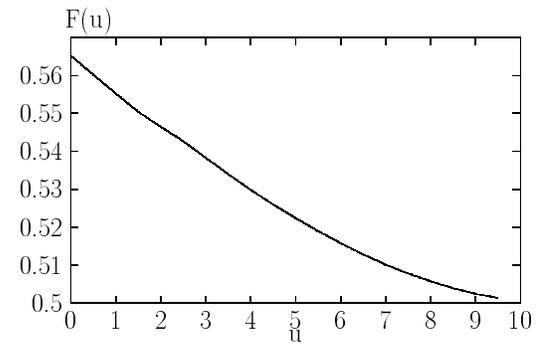
\item[Figure 3]  The evolution of the eccentricity for the {\it Tolman V-like%
} model with $\frac{M(0)}{A(0)}=0.217.$ Figures 3a through 3c correspond to
functions: $f(u)~=~cte=~1;$ $f(u)~=~1+u;$ and $f(u)~=~1+C\exp -\left( \frac{%
u-u_{trig}}\sigma \right) ^2$  , respectively. Figure 3d indicates the
evolution of the gravitational potential at the surface: $F(u)=1-\frac{2M(u)%
}{A(u)}$

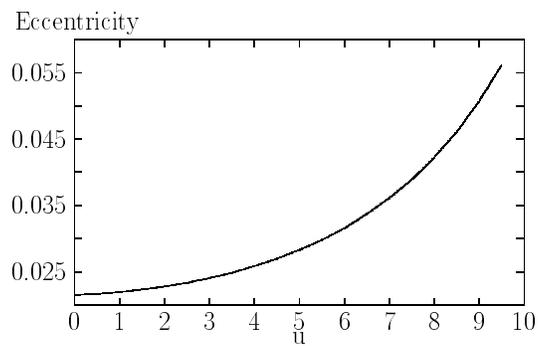
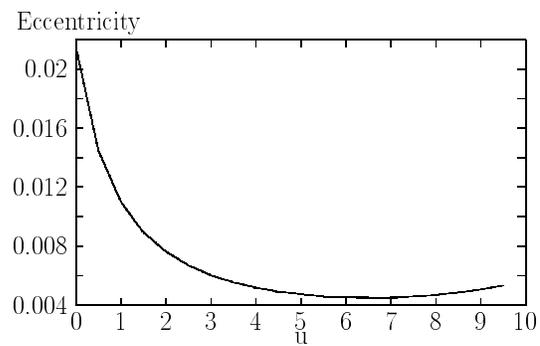
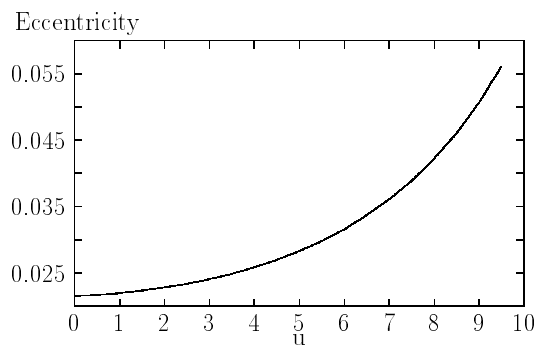
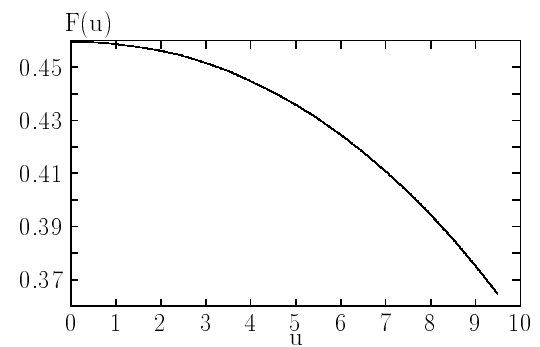
\item[Figure 4]  The evolution of the eccentricity for the {\it Tolman
VI-like} model with $\frac{M(0)}{A(0)}=0.270.$ Figures 4a through 4c
correspond to functions: $f(u)~=~cte=~1;$ $f(u)~=~1+u;$ and $f(u)~=~1+C\exp
-\left( \frac{u-u_{trig}}\sigma \right) ^2$  , respectively. Figure 4d
indicates the evolution of the gravitational potential at the surface: $%
F(u)=1-\frac{2M(u)}{A(u)}$
\end{description}

\end{document}